 \documentclass{emulateapj}

\usepackage{graphicx,natbib}
\usepackage{hyperref}
\shorttitle{V773~Cas, QS~Aql, and BR~Ind: eclipsing binaries as parts of multiple systems.}
\shortauthors{Zasche et al.}

\begin{document}

   \title{V773~Cas, QS~Aql, and BR~Ind: eclipsing binaries \\ as parts of multiple systems\thanks{Based on observations
   collected at the European Organisation for Astronomical Research in the Southern Hemisphere under ESO
   programmes 091.D-0122(A), 094.A-9029(D), 095.A-9032(A), and 096.A-9039(A), and also on the data from 2-m
   telescope at the Ond\v{r}ejov observatory in Czech Republic.}}

   \author{P. Zasche\altaffilmark{1}
          \and
          J. Jury\v{s}ek\altaffilmark{1,5}
          \and
          J. Nemravov\'a\altaffilmark{1}
          \and
          R. Uhla\v{r}\altaffilmark{2}
          \and
          P. Svoboda\altaffilmark{3}
          \and
          M. Wolf\altaffilmark{1}
          \and
          K. Ho\v{n}kov\'a\altaffilmark{4}
          \and
          M. Ma\v{s}ek\altaffilmark{5}
          \and
          M. Prouza\altaffilmark{5}
          \and \\
          J. \v{C}echura\altaffilmark{6}
          \and
          D. Kor\v{c}\'akov\'a\altaffilmark{1}
          \and
          M. \v{S}lechta\altaffilmark{6}
          }


 \affil{
  \altaffilmark{1} Astronomical Institute, Charles University in Prague, Faculty of Mathematics and Physics, CZ-180~00, Praha 8, \\
             V~Hole\v{s}ovi\v{c}k\'ach 2, Czech Republic\\
              \email{zasche@sirrah.troja.mff.cuni.cz}
  \altaffilmark{2}
             Private Observatory, Poho\v{r}\'{\i} 71, CZ-254 01, J\'{\i}lov\'e u Prahy, Czech Republic \\
  \altaffilmark{3}
             Private Observatory, V\'ypustky 5, CZ-614 00, Brno, Czech Republic \\
  \altaffilmark{4}
             Variable Star and Exoplanet Section of Czech Astronomical Society, Vset\'{\i}nsk\'a 941/78, CZ-757 01, Vala\v{s}sk\'e Mezi\v{r}\'{\i}\v{c}\'{\i}, Czech Republic  \\
  \altaffilmark{5}
             Institute of Physics, The Czech Academy of Sciences, Na Slovance 1999/2, CZ-182 21, Praha, Czech Republic \\
  \altaffilmark{6}
             Astronomical Institute, The Czech Academy of Sciences, CZ-251 65, Ond\v{r}ejov, Czech Republic \\
 }


\begin{abstract}
\noindent
   Eclipsing binaries still represent crucial objects for our understanding of the Universe. Especially those ones which are
   components of the multiple systems can help us solving the problem of their formation.
   The radial velocities together with the light curve analysis produced for the first time the precise physical
   parameters of the components of the multiple systems V773~Cas, QS~Aql, and BR~Ind. Their visual orbits were also
   analysed, resulted in slightly improved orbital elements.
   What is typical for all these systems is the fact that in all of them the most dominant source is
   the third distant component. The system V773 Cas consists of two similar G1-2V stars revolving on a
   circular orbit, while the more distant component is of A3V type. Additionally, the improved
   value of parallax was calculated to be of 17.6~mas. Analysis of QS Aql resulted in the
   following: inner eclipsing pair is composed of B6V and F1V stars, and the third component is of
   about B6 spectral type. The outer orbit has high eccentricity of about 0.95, and the observations
   near its upcoming periastron passage in between the years 2038-2040 are
   of high importance. Also the parallax of the system was derived to be of about 2.89~mas,
   moving the star much closer to the Sun than originally assumed.
   System BR~Ind was found to be a quadruple star consisting of two eclipsing K
   dwarfs orbiting around each other with the period of 1.786 days, while the distant component
   is a single-lined spectroscopic binary with the orbital period of about 6 days. Both pairs
   are moving around each other on its 148 yr orbit.
\end{abstract}

   \keywords{stars: binaries: eclipsing -- stars: binaries: spectroscopic -- stars: fundamental
   parameters -- stars: individual: V773 Cas, QS Aql, BR Ind}

\section{Introduction}

Eclipsing binaries and multiple systems play a crucial role in our understanding
of the Universe. The eclipsing binaries are being used for precise derivation of the stellar
parameters such as radii, masses or luminosities. On the other hand, the multiple systems play an
important role in our calibrations of models of star formation and evolution, because the presence
of triple, quadruple, or even higher order systems can serve as a very sensitive indicator in these
models and simulations. And finally, this distant component can play a crucial role during the
evolution of the system, it offers the possibility to study the role of the so-called Kozai cycles
with tidal friction (see e.g. \citealt{2001ApJ...562.1012E}) or try to detect a slow precession
of both inner and outer orbits.

Using the nowadays technique and large telescopes, automatic surveys and satellite observatories
the borders of the astrophysical front-line research is still moving to the more distant and more
faint targets. This is quite a logical process, but we have to be very careful when saying anything
about completeness of our knowledge of bright and close systems. As was already
stated in several recently published papers, also relatively bright targets among eclipsing
binaries located within one hundred parsecs distance from the Sun can bring us quite new surprising results
(see e.g. \citealt{2011A&A...532A..50M}, or \citealt{2016KsiTau}).

Therefore, we focused our effort on three rather seldom-investigated systems (namely V773~Cas,
QS~Aql, and BR~Ind) containing besides an inner eclipsing pair also a more distant third
component detected via interferometry and having the orbital periods from several decades to hundreds
of years (hence their ratio of periods is very large). Moreover, all of these stars show an
Algol-like light curve and their spectroscopic as well as photometric study is still missing yet.
As was already published earlier e.g. by \cite{2009AJ....138..664Z} a list of such systems with
eclipsing components among visual doubles, where both the inner and outer orbits are known is
still very limited to only several dozens on the whole sky.

The statistics of the triple and multiple systems is still rather limited yet, but what can surely
be said is that there is a lack of systems with higher-mass tertiary among the triple stars. This
was shown e.g. by \cite{2008MNRAS.389..925T} on spectroscopic triple stars or by
\cite{2016MNRAS.455.4136B} on the Kepler eclipsing binaries. Only a small fraction of systems have
the more massive tertiary than the eclipsing pair itself. And here comes our contribution to the
topic.

\section{The data}

The spectroscopy was obtained in two observatories. Most of the data
points for these systems came from the Ond\v{r}ejov observatory and its 2-meter telescope
(resolution R $\sim$ 12500). Additionally, the data for BR~Ind and some of the data for QS~Aql were
obtained with the FEROS instrument mounted on 2.2-meter MPG telescope located in La Silla
Observatory in Chile (R $\sim$ 48000). The individual exposing times were chosen according to the
quality of the particular night and specifications of the instrument to achieve the $S/N$ ratio
between several dozens to a few hundreds.

The original FEROS data were reduced using the standard routines. The final radial velocities
(hereafter RV) used for the analysis were derived via a technique comparing both
the direct and flipped profile of the spectral lines manually on the computer screen to
achieve the best match, using program SPEFO (\citealt{1996A&A...309..521H},
\citealt{1996ASPC..101..187S}) on several absorption lines in the measured spectral region (usually
\emph{Fe}, \emph{Ca}, or \emph{Si} lines). The derived radial velocities are given in Tables below
in Appendix section.

The photometry for these three systems were collected during the time span from 2008 to 2016. However, some
of the older data used only for the minima times were already published earlier, but the complete
light curves (hereafter LC) are being published here for the first time. All of the data were
obtained in the Johnson-Cousins photometric system \cite{1990PASP..102.1181B}, particularly the
system V773~Cas in $BVR$, while both the systems QS~Aql and BR~Ind in $BVRI$ filters.

Owing to the relatively high brightness of the targets, only rather small telescopes were used for
these photometric observations. The system V773~Cas has been observed (by one of the authors: PS)
with the 34-mm refractor at the private observatory in Brno, Czech Republic, using the SBIG ST-7XME
CCD camera. The star QS~Aql was monitored with the similar instrument at the private observatory
(by one of the authors: RU) in J\'{\i}lov\'e u Prahy, Czech Republic, using a G2-0402 CCD camera.
On the other hand, the only very southern star BR~Ind was observed with the FRAM telescope
\citep{2010AdAst2010E..31P}, installed and operated at the Pierre Auger Observatory at Malarg\"ue,
Argentina. For observations only a small Nikkor lens with 107 mm diameter and a CCD camera of
G4-16000 type was used (which is mounted on 30-cm FRAM telescope itself). All the measurements were
processed by the software C-MUNIPACK\footnote{See http://c-munipack.sourceforge.net/} which is
based on aperture photometry and uses the standard DAOPHOT routines \citep{1993ASPC...52..173T}.

\section{The analysis}

The whole work is based on classical techniques of using the photometry and spectroscopy together
with the analysis of the positional measurements of the visual double on the sky obtained during
much longer time span (more than a century). Combining these methods together one can not only
obtain the reliable orbital and stellar parameters, but also the structure of the system and its long-term evolution.
The advantage is also the fact that having the complete information about the masses, inclinations,
periods, etc. we can also fill in still rather incomplete statistics of the triple and quadruple
systems, which is compared to models of formation of binaries and multiple systems
\citep{2008MNRAS.389..925T}.

At first, the visual orbit based on already published interferometric data was analysed.
However, orbits of systems analysed within this study were published quite recently.
Hence, our new re-calculations led to only slight improvements of the fits. The data were
downloaded from the already published papers and the Washington Double Star Catalogue
(hereafter WDS\footnote{http://ad.usno.navy.mil/wds/}, \citealt{WDS}). The orbits were calculated following
the paper \cite{2007AN....328..928Z}, but the coverage of the orbits was usually not perfect and
only parts of the long orbits are covered with data nowadays.

Both the photometry and spectroscopy were studied in the standard manner. The obtained photometric
data and the radial velocities were analysed by the program {\sc Phoebe}
\citep{2005ApJ...628..426P}, which is using the classical Wilson-Devinney algorithm
(\cite{1971ApJ...166..605W} and its later modifications) and allows us to fit the relevant
parameters of the eclipsing components and their relative orbit. For the modelling, we used several
assumptions. At first, the primary temperature was set to the value corresponding to the particular
spectral type (see e.g. calibrations by \cite{2013ApJS..208....9P} and the updated web
site\footnote{\tiny{http://www.pas.rochester.edu/$\sim$emamajek/EEM$\_$dwarf$\_$UBVIJHK$\_$colors$\_$Teff.txt}}).
The limb-darkening coefficients were obtained through interpolation in tables by
\citet{1993AJ....106.2096V}. The albedo coefficients $A_i$, and the gravity darkening coefficients
$g_i$ 
were fixed at their suggested values according to the temperatures of the components. As all
studied eclipsing binaries are members of multiple systems third light from the remaining
components was taken into account.

And finally, if we have the LC+RV solution and both the eclipsing binary masses are known, we can
proceed to the combined analysis of the visual orbit together with the period changes of the
eclipsing pair. The method itself was introduced in
\cite{2007AN....328..928Z}, and its usage was presented e.g. in \cite{2012A&A...542A..78Z}, or
\cite{2014AcA....64..125Z}. The most crucial for the whole analysis seems to be the quality of the
input observations in both methods and the data coverage of the long orbit. This is usually
problematic in these cases where the third-body orbit is too long and we have only small fraction
of the orbit covered. On the other hand, if we have a good data coverage in both methods, we can
even calculate independently the distance to the system.

\section{V773 Cas}

The first system in our sample of stars is the northern-hemisphere V773~Cas (= HIP~8115, HD~10543,
$V_{max}$ = 6.18~mag), an eclipsing binary discovered on the basis of Hipparcos data
(\citealt{HIP}, and \citealt{1999IBVS.4659....1K}). However, many years before the discovery of its
photometric variability was the star recognized as a visual binary. Its most recent orbital
solution is that one published by \cite{2009AJ....138..813H}, who derived its period to be of about
193~yr, the orbital eccentricity of about 0.77, and the semimajor axis of about
0.9$^{\prime\prime}$. The spectral type was usually stated as A3V \citep{1978BICDS..15..121J} or
A2V \citep{1968RGOB..135..385P}. However, as noted by \cite{2010SerAJ.180...71C}, there arises a
large discrepancy between the astrophysical and dynamical total mass of the system. From its
spectral types the total mass should be of about 3~M$_\odot$, while from the orbital solution
arises that the $M_{dyn} = 11.9$~M$_\odot$. This strange discrepancy originally led to our interest
about the star.

 \begin{figure}
   \centering
   \includegraphics[width=0.48\textwidth]{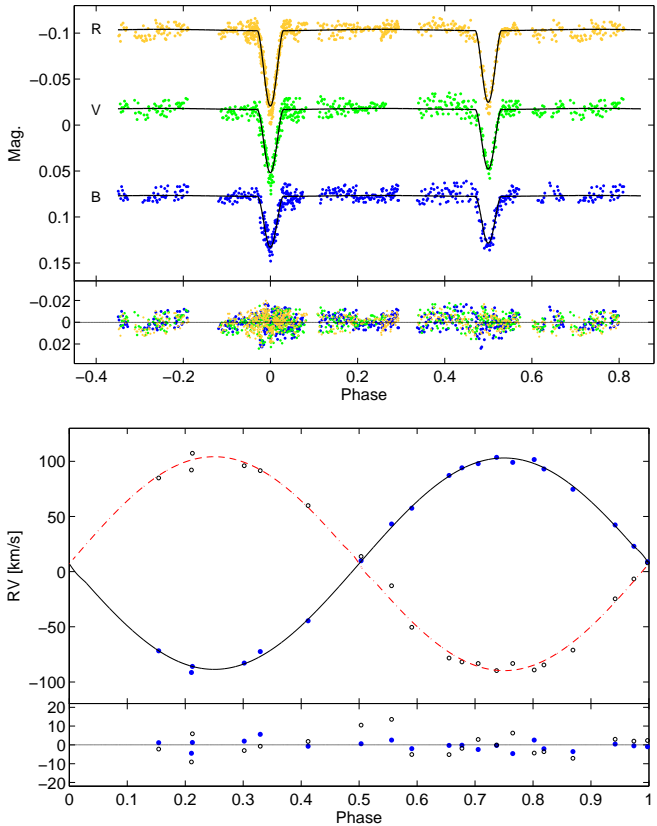}
   \caption{Light and radial velocity curve fits of V773~Cas based on the {\sc Phoebe} fitting.}
   \label{FigLCRV_V773Cas}
  \end{figure}

First of all, we found out that for V773~Cas three new interferometric measurements were obtained
since the last visual orbit calculation was published by \cite{2009AJ....138..813H}. We added these
three data points and re-ran the fitting procedure again, resulted only in slightly different
orbital solution. 
However, our solution would not be complete enough if we did not try to incorporate also the
photometric monitoring and our results from the LC+RV analysis.

\begin{table}
  \caption{The parameters from the LC+RV fitting of V773 Cas.}
  \label{V773CasLCRVparam}
  \centering
  \begin{tabular}{c c c c}
\hline \hline
Parameter       & Primary & Secondary & Tertiary \\
 \hline
 $HJD_0$ & \multicolumn{2}{c}{2448500.9209 $\pm$ 0.0003 } & -- \\
 $P$ [d] &  \multicolumn{2}{c}{2.587332 $\pm$ 0.000002} & -- \\
 $a$ [R$_\odot$] &  \multicolumn{2}{c}{9.96 $\pm$ 0.06} & -- \\
 $v_\gamma$ [km/s] & \multicolumn{2}{c}{7.11 $\pm$ 0.30 } & -- \\
 $q = M_2/M_1$ &  \multicolumn{2}{c}{1.00 $\pm$ 0.05 } & -- \\
 $i$ [deg]& \multicolumn{2}{c}{84.7 $\pm$ 2.2 } & -- \\
 $K$ [km/s] &  97.1 $\pm$ 0.9 & 97.0 $\pm$ 1.6  & -- \\
 $T$ [K] & 5900 (fixed) & 5842 $\pm 50$ & -- \\
 $M$ [M$_\odot$] & 0.99 $\pm$ 0.03 & 0.99 $\pm$ 0.04 & -- \\
 $R$ [R$_\odot$] & 1.05 $\pm$ 0.05 & 1.05 $\pm$ 0.05 & -- \\
 $M_{bol}$ [mag] & 4.55 $\pm$ 0.10 & 4.58 $\pm$ 0.10 & -- \\
 $L_B [\%]$ & 10.0 $\pm$ 0.9 &  9.5 $\pm$ 0.9 & 80.5 $\pm$ 0.9 \\
 $L_V [\%]$ & 12.5 $\pm$ 0.7 & 12.1 $\pm$ 0.7 & 75.4 $\pm$ 0.6 \\
 $L_R [\%]$ & 15.0 $\pm$ 0.6 & 14.6 $\pm$ 0.6 & 70.4 $\pm$ 0.5 \\
 \hline
\end{tabular}
\end{table}

The primary temperature was kept fixed at a value of 5900~K, which resulted from our spectral
estimations and also from the primary mass. The results of our LC+RV solution are shown in Figure
\ref{FigLCRV_V773Cas}, where one can see the light curve in $BVR$ filters together with the RV
curve based on the Ond\v{r}ejov data. In total, there were obtained 20 spectrograms and 15 nights
of photometry. The radial velocities were mostly derived from the \emph{Ca}\,I and \emph{Fe}\,I
lines. The parameters of the least-squares fit are given in Table \ref{V773CasLCRVparam}. The
system is well-detached and both eclipsing components are probably of about G1-2V spectral type,
hence should be considered as solar analogues. The spectral lines of both components seem to be
very similar to each other, while some of the lines which remained at almost fixed position (at
about 5~km$\cdot$s$^{-1}$) seem to originate from the third distant body, whose movement is
negligible over the time span of the observed spectra. However, as one can see from the relatively
high value of the third light as resulted from the LC solution, the third body is the brightest
member of the system and it is probably responsible for the spectral classification of V773~Cas as
A2-3 in the past.

 \begin{figure}
   \centering
   \includegraphics[width=0.48\textwidth]{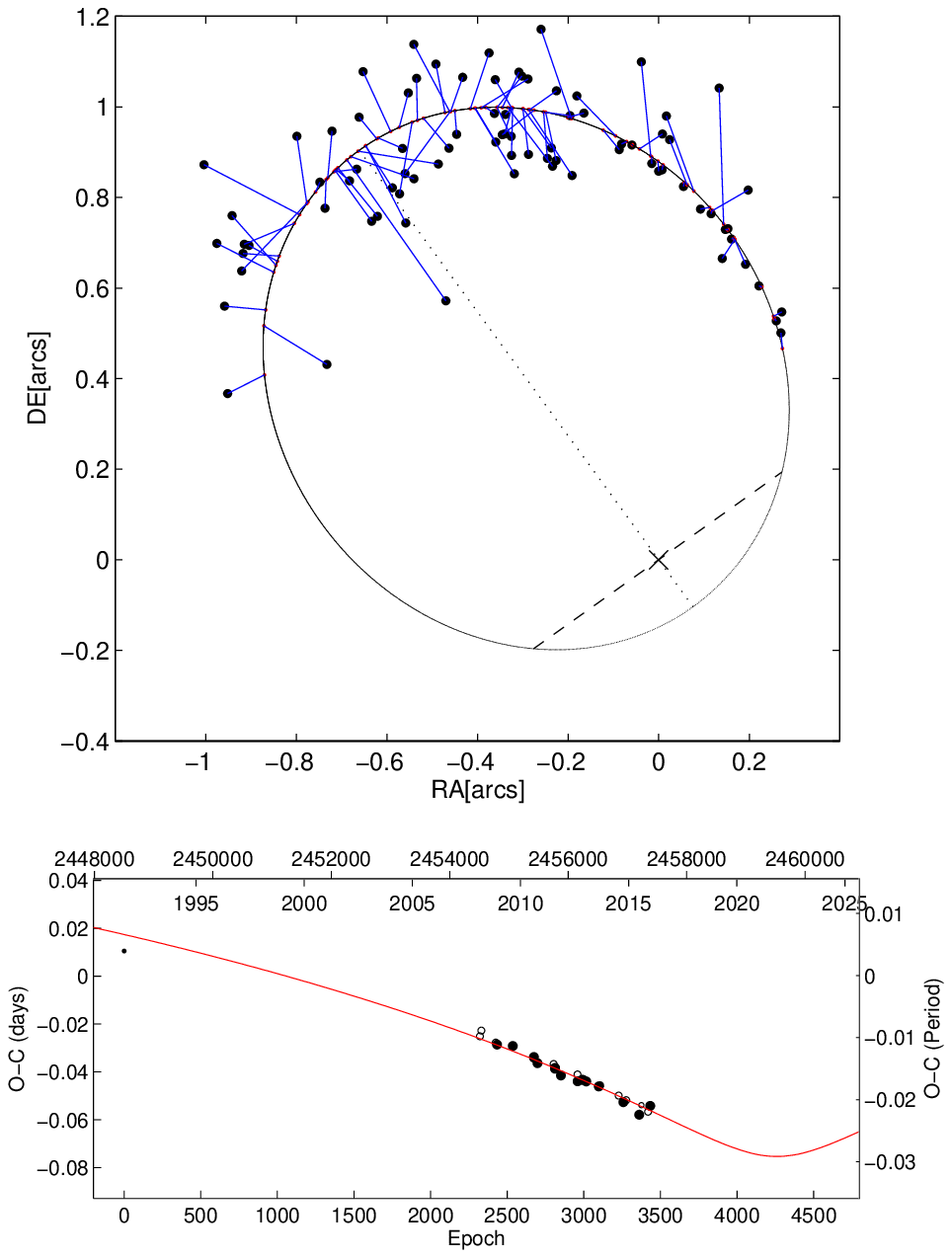}
   \caption{Upper plot: Orbit of V773 Cas on the sky. The individual observations are connected with their
   theoretical positions on the orbit, while the dotted line stands for the line of the apsides and
   the dashed line stands for the line of the nodes. Bottom plot: $O-C$ diagram of V773 Cas as resulted from our combined analysis of period changes
   together with the visual orbit. Best-fitting solution is plotted as a solid line, while the individual
   observations are denoted as filled (primary) and open (secondary) circles.}
   \label{FigV773Cas_OC_orbit}
  \end{figure}

The above-mentioned solution was derived using the {\sc Phoebe} code and the LC+RV fitting.
However, we also tried a different approach to the problem. Using the available 20 spectrograms we
applied the code {\sc
Pyterpol}\footnote{https://github.com/chrysante87/pyterpol/wiki}\citep{2016KsiTau}. This program
determines kinematic and radiative properties of binary components through comparison of observed
spectra to synthetic ones obtained through interpolation in pre-calculated grids of synthetic
spectra (\citealt{2012A&A...544A.126D} and \citealt{2010A&A...516A..13P}). Using this approach, we
obtained the solution presented in Table \ref{Table_V773Cas_Pyterpol}. Such result is in very good
agreement with the {\sc Phoebe} solution presented above as well as with the observed magnitude
difference between the two visual components (of about 1-2 mag from the WDS catalogue).

\begin{table}
  \caption{Parameters of V773 Cas using \sc{Pyterpol}.}
  \label{Table_V773Cas_Pyterpol}
  \centering
  \begin{tabular}{c c c c}
\hline \hline
Parameter         & Primary & Secondary & Tertiary \\
 \hline
 $T$ [K]          &  5933 $\pm$ 131  &  5693 $\pm$ 161  &  8522 $\pm$ 38  \\
 $v \sin i$ [km/s]& 32.17 $\pm$ 2.32 & 49.10 $\pm$ 7.46 & 84.55 $\pm$ 1.42\\
 $L$ [\%]         & 13.1  $\pm$ 0.8  & 11.1  $\pm$ 0.5  & 75.8  $\pm$ 0.6 \\ \hline
\end{tabular}
\end{table}

The linear ephemerides written in Table \ref{V773CasLCRVparam} are the best suitable elements for
future prospective observations of V773~Cas in the upcoming years. However, these elements will
change significantly due to the orbital motion of the eclipsing pair around a common barycenter
with the third component. The most significant change of the orbital elements of the inner pair
will take place near the periastron passage which will occur in 2021. We plotted the predicted
period variation of V773~Cas eclipsing pair in the $O-C$ diagram in the Fig.
\ref{FigV773Cas_OC_orbit}. This diagram was constructed in agreement with the visual orbit of the
double as resulted from our combined analysis. The orbit of the third component is given in Fig.
\ref{FigV773Cas_OC_orbit} and the parameters of such a fit are given in Table
\ref{Table_AstrOrbitV773Cas}. The list of minima times used for the analysis are given below in
Appendix section.

\begin{table}
  \caption{The orbital parameters of V773 Cas.}
  \label{Table_AstrOrbitV773Cas}
  \centering
  \begin{tabular}{c c c}
\hline \hline
Parameter       & Our solution & \cite{2009AJ....138..813H} \\
 \hline
 $p_3$ [yr]     & 184.9  $\pm$  2.7   & 193.17 $\pm$ 6.23 \\
 $T_0$ [yr]     & 2021.8 $\pm$  2.1   & 2022.39 $\pm$ 0.78 \\
 $e$            & 0.794  $\pm$  0.050 & 0.773 $\pm$ 0.016 \\
 $a$ [arcsec]   & 0.911  $\pm$  0.065 & 0.899 $\pm$ 0.012 \\
 $i$ [deg]      & 133.3  $\pm$  2.6   & 134.8 $\pm$ 3.8 \\
 $\Omega$ [deg] & 125.4  $\pm$  4.3   & 128.8 $\pm$ 4.6 \\
 $\omega$ [deg] & 269.5  $\pm$  8.5   & 270.2 $\pm$ 7.4 \\
 \hline
\end{tabular}
\end{table}

The problem is that for achieving such a self-consistent solution, we cannot use the Hipparcos
parallax as an input parameter. The spectral classification of about A3V for the third component
comes not only from the already published papers, but also from our findings about the spectra (the
lines indicate a spectral type of about A3), as well as from the photometric indices of the third
body as resulted from the LC+RV solution. Hence, the total mass of the three components should be
of about 4~M$_\odot$, which is in contradiction with the computed mass using the Hipparcos parallax
$\pi_{HIP}=11.77 \pm 0.67$~mas. Hence, the parallax needed for our combined solution to be
self-consistent one needs the value of about $\pi_{new}=17.6 \pm 1.5$~mas, but its uncertainty is
still rather high because it is based only on a mass estimation. Such a situation is nothing novel,
because there was already shown that the Hipparcos data sometimes produce spurious results for the
close double stars, see e.g. \cite{2008AJ....136..890D}.

\section{QS Aql}

Second eclipsing system analysed in the present study is QS~Aql (=HIP~96840, HD~185936, $V_{max}$ =
6.01~mag), which is the brightest one and also the most massive one among the studied systems.
\cite{1978BICDS..15..121J} classified its spectral type as B5V, while the others like
\cite{1923AnHar..98....1C} or \cite{1991PBeiO..17...59L} published its type as B3. Moreover, its
variability was first detected by \cite{Millman1928}, but its eclipsing nature was confirmed by
\cite{Guth1931}, who also gave its correct orbital period of about 2.5~days. Some 40 years later,
\cite{Knipe1971} discovered a rapid period change, which occurred at about 1964 (his suggestion)
and was caused by the periastron passage in the wide orbit around the barycenter. The period change
was so rapid that the eccentricity of the wide orbit must be very high. On the other hand, the
first astrometric observations are more than 80~years old, but their accuracy is questionable due
to small angular separation of the components. Many reliable speckle interferometric
observations were obtained since 1976. The most recent orbital solution was computed by
\cite{2007AJ....133.1209D}, who derived its orbital period to be of 61.72~yr and surprisingly high
eccentricity of about 0.966. The total mass of the system was estimated to be of about
20~M$_\odot$, but with rather high uncertainty. \cite{Mayer2004} noted that the combined analysis
of period changes together with the visual orbit is still problematic due to poor coverage of data
by both methods.

We started the photometric monitoring of this interesting system in 2007, while the new
spectroscopy was collected from 2012. Since the last calculation of its visual orbit by
\cite{2007AJ....133.1209D} there was published one new observation of the visual double. The system
is known as a single-lined spectroscopic binary, while the secondary as well as the tertiary
component lines were not detected in the spectra. For the discussion about the individual RV
solutions see below.

 \begin{figure}
   \centering
   \includegraphics[width=0.48\textwidth]{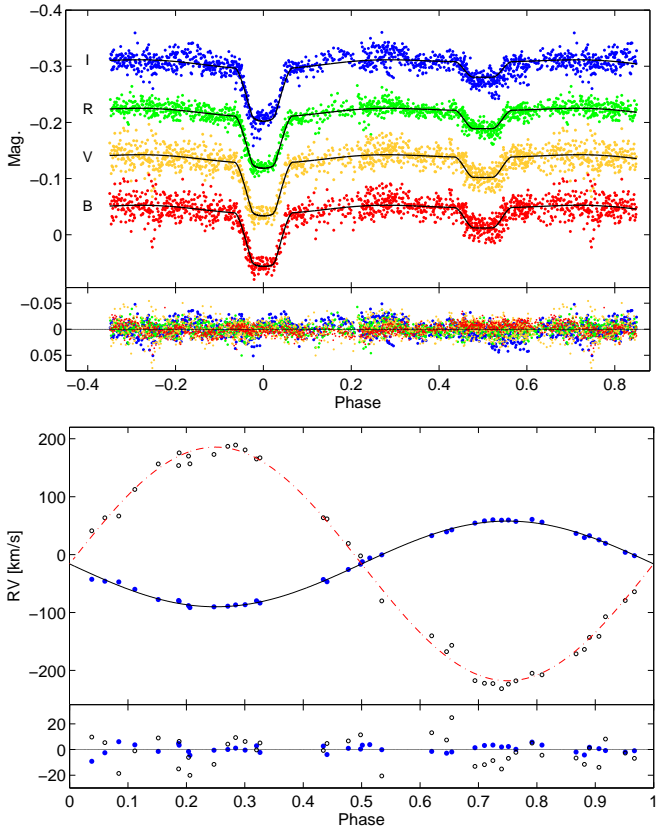}
   \caption{Light and radial velocity curve fits of QS~Aql based on the {\sc Phoebe} fitting.}
   \label{FigLCRV_QSAql}
  \end{figure}

\begin{table}
  \caption{The parameters from the LC+RV fitting of QS Aql.}
  \label{QSAqlLCRVparam}
  \centering
  \begin{tabular}{c c c c}
\hline \hline
Parameter       & Primary & Secondary & Tertiary \\
 \hline
 $HJD_0$ & \multicolumn{2}{c}{2440443.5442 $\pm$ 0.0015 } & -- \\
 $P$ [d] &  \multicolumn{2}{c}{2.5132987 $\pm$ 0.0000075 } & -- \\
 $a$ [R$_\odot$] &  \multicolumn{2}{c}{13.78 $\pm$ 0.11 } & -- \\
 $v_\gamma$ [km/s] & \multicolumn{2}{c}{-16.13 $\pm$ 0.62 } & -- \\
 $q = M_2/M_1$ &  \multicolumn{2}{c}{0.37 $\pm$ 0.02 } & -- \\  
 $i$ [deg]& \multicolumn{2}{c}{83.6 $\pm$ 1.3 } & -- \\
 $K$ [km/s] &  73.98 $\pm$ 0.33 & 201.76 $\pm$ 2.09 & -- \\
 $T$ [K] & 14500 (fixed) & 7910 $\pm$ 78 & -- \\
 $M$ [M$_\odot$] &  4.07 $\pm$ 0.09 & 1.49 $\pm$ 0.05 & -- \\
 $R$ [R$_\odot$] &  4.08 $\pm$ 0.15 & 1.65 $\pm$ 0.20 & -- \\
 $M_{bol}$ [mag] & -2.31 $\pm$ 0.18 & 2.29 $\pm$ 0.14 & -- \\
 $L_B [\%]$ & 47.6 $\pm$ 2.9 &  1.4 $\pm$ 0.4 & 51.0 $\pm$ 3.1 \\
 $L_V [\%]$ & 47.4 $\pm$ 1.2 &  2.0 $\pm$ 0.3 & 50.6 $\pm$ 1.4 \\
 $L_R [\%]$ & 49.2 $\pm$ 3.4 &  2.3 $\pm$ 0.2 & 48.5 $\pm$ 3.5 \\
 $L_I [\%]$ & 48.7 $\pm$ 2.3 &  2.7 $\pm$ 0.2 & 48.6 $\pm$ 2.5 \\
 \hline
\end{tabular}
\end{table}

The light curve analysis was carried out using the data obtained in $BVRI$ filters in the Czech
Republic in 2009 and 2010. The results are shown in Fig. \ref{FigLCRV_QSAql}, while the parameters
corresponding to the best-fitting synthetic light curve are given in Table \ref{QSAqlLCRVparam}. In
Fig. \ref{FigLCRV_QSAql} you can also see some small variability on the residuals after subtraction
of the light curve. However, these deviations are only caused by worse photometric conditions
during some of the nights and our decision not to remove the outlying points on the light curves.
The results of the RV fitting are shown in Fig. \ref{FigLCRV_QSAql}, where one can see that the
secondary velocities were also derived, but are affected by much larger errors than primary ones.

 \begin{figure}
   \centering
   \includegraphics[width=0.48\textwidth]{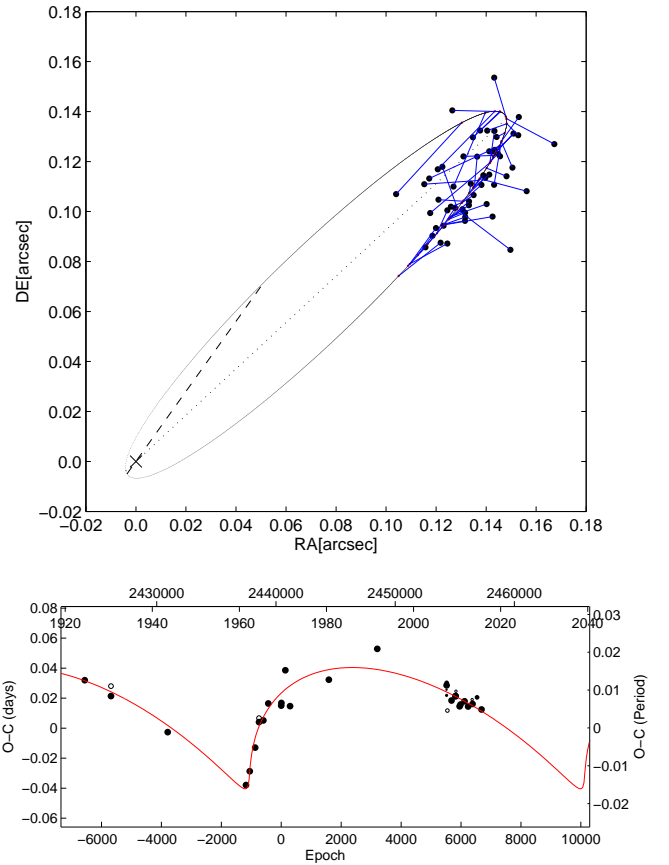}
   \caption{Orbit of QS Aql on the sky, and the $O-C$ diagram. See Fig. \ref{FigV773Cas_OC_orbit} for description.}
   \label{FigQSAql_OC_orbit}
  \end{figure}

 \begin{figure}
   \centering
   \includegraphics[width=0.48\textwidth]{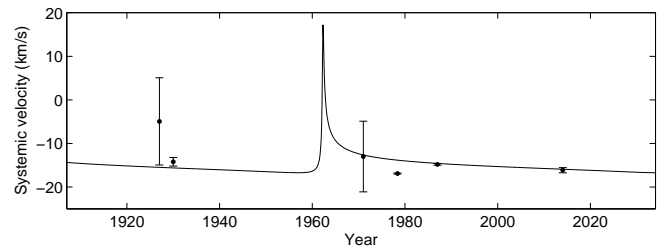}
   \caption{Gamma velocities of QS Aql as based on individual published RV solutions.}
   \label{FigRVvar_QSAql}
  \end{figure}

As one can see from the parameters given in Table \ref{QSAqlLCRVparam}, the eclipsing components
are rather different, but the most luminous one seems to be the third distant member. What is
rather surprising is the fact that the amplitude of the RV variations for the primary component
resulted in about 74~km/s, while the authors of the previous studies gave the $K_1$ value in
between 40.8~km/s \citep{1990ApJS...74..551A} and 58.4~km/s \citep{1987BAAS...19..709H}. The other
solutions gave also rather low values of $K_1$ about 47.3~km/s, see \cite{Hill1931} and
\cite{Lucy1971}. Explanation of this discrepancy probably comes from the fact that the lines are
very broad and blended together with the third (dominant) component, which remains on almost the
same position over the whole time interval. Thanks to this reason the lines are rather asymmetric
and the previous authors probably measured the wide wings of the lines instead of the cores. If we
measure the wings, the amplitude is really lower. However, thanks to the high dispersion FEROS
spectra we were able to confidentially identify both the eclipsing components for the first time
(hence SB1 becomes SB2) and derive the amplitude $K_1$ with higher conclusiveness.

If we tried to obtain a combined solution of the visual orbit plus the period variation of the
eclipsing pair, we found out that the analysis is still rather problematic. For the trustworthy fit
of the visual orbit only the data obtained since 1976 were taken into account. On the other hand,
the older times of minima were also used because these define the rapid period change near the year
1960 quite well. For the results of our fitting see Fig. \ref{FigQSAql_OC_orbit}. The parameters of
our solution are presented in Table \ref{Table_combined_QSAql}.

As a by-product of the fitting we also plotted the gamma velocity changes in agreement with the
third body orbit and plotted there also the individual gamma velocities as resulted from individual
studies as published during the last century. Our result is shown in Fig. \ref{FigRVvar_QSAql}.
Unfortunately, some individual systemic velocities are affected by large uncertainties and the
predicted RV variation is rather flat during most of the $p_3$ period.

\begin{table}
  \caption{The parameters of our combined solution for QS Aql.}
  \label{Table_combined_QSAql}
  \centering
  \begin{tabular}{c c c}
\hline \hline
Parameter       & Value   \\
 \hline
 $p_3$ [yr]     &  77.0  $\pm$ 4.3   \\
 $T_0$ [yr]     & 1962.3 $\pm$ 2.3   \\
 $e$            & 0.947  $\pm$ 0.038 \\
 $a$ [arcsec]   & 0.111  $\pm$ 0.045 \\
 $i$ [deg]      & 61.2   $\pm$ 3.6   \\
 $\Omega$ [deg] & 144.5  $\pm$ 5.1   \\
 $\omega$ [deg] & 336.8  $\pm$ 4.7   \\ \hline
 $M_3$ [M$_\odot$]& 4.04 $\pm$ 0.86  \\
 $\pi$ [mas]    & 2.89   $\pm$ 0.55  \\
 \hline
\end{tabular}
\end{table}

However, a discussion of this combined solution is necessary. The presented final fits are still
rather preliminary and especially the astrometry suffers from many deviating points. This is
probably caused by rather large eccentricity and the inclined orbit. Several much more reliable
observations would be very useful in the upcoming years. In the paper \cite{Heintze1989} the
authors discussed the spectroscopic orbit by \cite{1987BAAS...19..709H} and concluded that the
third light should be of about 1.2 times larger than the combined light of the eclipsing pair.
Their conclusion would imply tertiary mass 4.3~M$_\odot$ and spectral type of B5-6\,V. However,
this assumption is contradicted by the last interferometric observations, which indicate
that both visual components are of similar brightness (i.e.
the combined light from the eclipsing pair roughly the same as the third star), which is also in
agreement with our new LC+RV solution. From this information we can derive that the third component
is probably of about the same spectral type as the primary, i.e. B6V with the mass of about
4~M$_\odot$. Exactly the same result was obtained from our combined analysis of period changes
and the visual orbit, see Table~\ref{Table_combined_QSAql}.

Thanks to the relatively well-derived amplitudes of both phenomena (semimajor axis for the visual
orbit as well as the semiamplitude of the period variations in the $O-C$ diagram) we also tried
to determine independently distance to the system. Quite interesting is the fact how the
parallax of QS~Aql has changed from the original Hipparcos value 1.98 $\pm$ 0.82~mas \citep{HIP},
while the new one was recalculated to 0.49 $\pm$ 0.62~mas \citep{2007A&A...474..653V}. On the other
hand, \cite{Docobo2006} presented two different possible values of parallax based on two different
methods and the Hipparcos parallax, namely 1.8~mas and 3.1~mas. No other parallax estimation has
been found in the literature. However, from our solution there resulted that the parallax value
have to be larger than the Hipparcos ones due to the amplitudes in both methods, of about 2.89~mas.
The future space missions like GAIA \citep{2001A&A...369..339P} would solve this problem, however
the star is too bright and close to the bright limit of the satellite.

\section{BR Ind}

The last system in our analysis is rather neglected and only seldom-investigated BR~Ind (=
HIP~104604, HD 201427, $V_{max}$ = 7.07~mag). Its photometric variability was discovered on the
basis of the Hipparcos data \citep{HIP}, giving the orbital period of 0.89277~days (a short note
about its possible double value was also added there). Its spectral type F8V was published by
\cite{1978mcts.book.....H}, however it is not clear to which component this classification belongs.
The star is also known as a visual double, having the time span of the positional observations of
more than 100~yrs. The most recent orbital solution was published by \cite{Seymour2002}, who
derived the period of 167~yr, semimajor axis of 0.894$^{\prime\prime}$ and eccentricity of 0.521.
However, since than five new observations were obtained and the orbit should be recalculated.

We collected the available photometry of BR Ind trying to find out which of the orbital periods is
the correct one (0.89 or 1.78 days). However, the photometry from the surveys like ASAS
\citep{ASAS} or Pi of the sky \citep{2005NewA...10..409B} were not able to distinguish between
these two periods. Therefore, the photometry for BR~Ind in $BVRI$ filters was obtained in 2014
and 2015 using the FRAM telescope. Thanks to these data we finally confirmed that the correct orbital
period of the inner pair is really double, i.e. of about 1.786~days.

On the other hand, we also obtained the spectroscopy of BR Ind with the FEROS spectrograph in La
Silla. However, after four nights (and obtaining 27 \'echelle spectra) of observations we were not
able to detect the 1.8-days period on the most prominent lines. Instead, the lines follow some
longer periodicity of several days. Hence, we applied for more observing time using the Tycho Brahe
proposals for the 2.2-meter MPG telescope and the FEROS instrument again. During four consecutive
seasons we obtained 14 more spectra of BR~Ind leading finally to the solution.

 \begin{figure}
   \centering
   \includegraphics[width=0.48\textwidth]{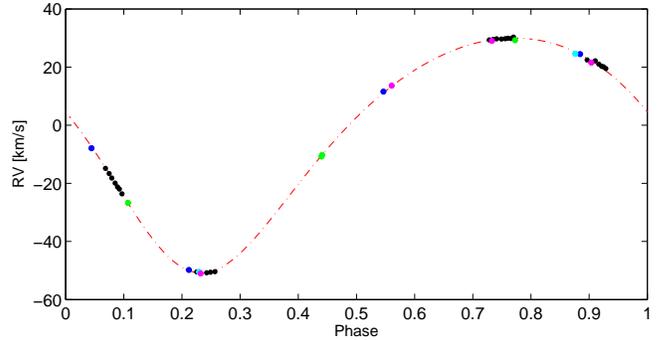}
   \caption{Radial velocity curve of the B pair of BR~Ind as obtained from the FEROS data. The black
   dots stand for the data from fall 2013, blue ones are from fall 2014, green dots denote the data
   from spring 2015, the cyan ones are the observations coming from fall 2015, while the magenta ones
   are from spring 2016.}
   \label{FigRV_BRIndB}
  \end{figure}

 \begin{figure}
   \centering
   \includegraphics[width=0.48\textwidth]{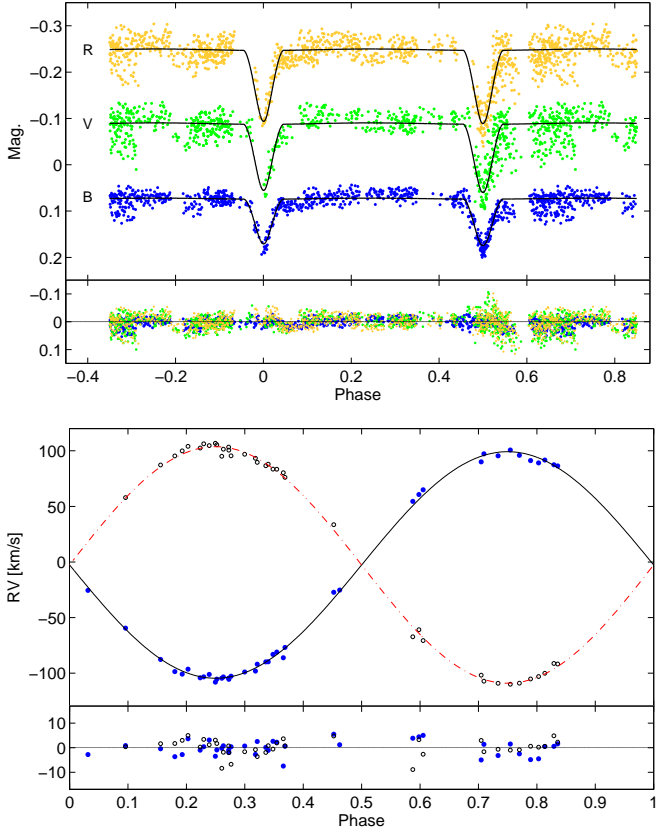}
   \caption{Light and radial velocity curve fits of the inner pair of BR~Ind based on the {\sc Phoebe} fitting.}
   \label{FigLCRV_BRInd}
  \end{figure}

  \begin{figure}
   \hspace{8mm}
   \includegraphics[width=0.28\textwidth]{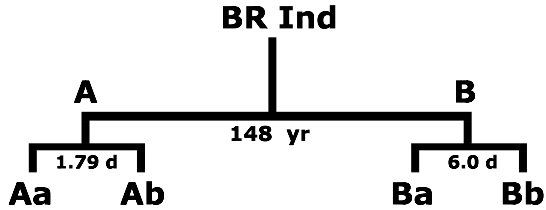}
   \caption{Structure of BR~Ind as resulted from our analysis.}
   \label{Fig_BRInd_structure}
  \end{figure}

The most prominent lines which were also used to derive the radial velocities were the $Fe$ and
$Ca$ lines. These were analysed leading to the detection of a 6-days variation. Only much weaker
lines were detected as the lines coming from the primary and secondary components of the eclipsing
pair and following the 1.786-days variability. Hence, for the subsequent analysis we denoted the
6-days orbit as the ''B'' component, while the eclipsing pair is always designated as ''A''. The
results of our RV fitting are plotted in Figs. \ref{FigRV_BRIndB} and \ref{FigLCRV_BRInd}.
Resulting parameters of the pair B are given in Table \ref{BRIndLCRVparam}. As one can see, the
orbit is only slightly eccentric. Structure of BR~Ind is plotted in Fig. \ref{Fig_BRInd_structure}.

\begin{figure}
   \centering
   \includegraphics[width=0.48\textwidth]{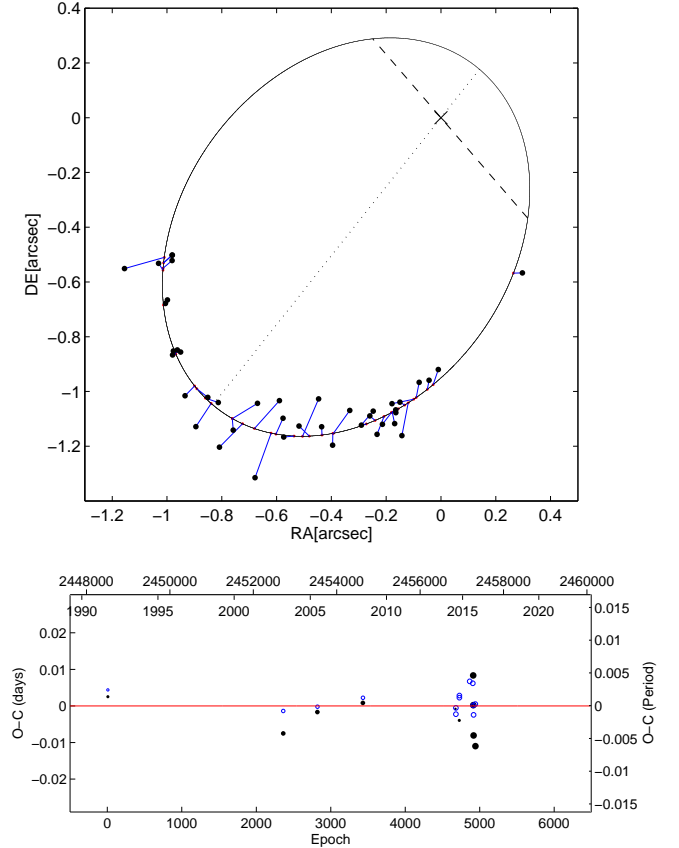}
   \caption{Orbit of BR Ind on the sky, and the $O-C$ diagram. See Fig. \ref{FigV773Cas_OC_orbit} for description.}
   \label{FigBRInd_OC_orbit}
  \end{figure}

The light curve fitting was carried out together with the radial velocity analysis in {\sc Phoebe}.
The results are plotted in Fig. \ref{FigLCRV_BRInd}, while the parameters are given in Table
\ref{BRIndLCRVparam}. We can see that both components are very similar to each other (therefore the
problems with 0.89 versus 1.78-days period). Both gamma velocities of the eclipsing pair as well as
of the B pair are similar to each other and close to zero. This would indicate that the orbit is
very close to face-on orientation, which was confirmed via the fitting of the astrometry (see
below). We also collected the available photometry for deriving the times of the eclipses and
constructed the $O-C$ diagram plotted in Fig. \ref{FigBRInd_OC_orbit}. No visible variation can be
seen there during these approximately 25 years of observations. This would also indicate that the
period of the visual pair is rather long or the orbit is nearly face-on.

\begin{table}
  \caption{The parameters from the LC+RV fitting of BR Ind.}
  \label{BRIndLCRVparam}
  \centering
  \scalebox{0.83}{
  \begin{tabular}{c c c c}
\hline \hline
                  &\multicolumn{2}{c}{Pair A}                   & Pair B \\
Parameter         & Primary   & Secondary                       &  \\
 \hline
 $HJD_0$          & \multicolumn{2}{c}{2448500.4755 $\pm$ 0.0002} & 2456563.186 $\pm$ 0.052 \\
 $P$ [d]          & \multicolumn{2}{c}{1.7855618 $\pm$ 0.0000015} & 6.0009949 $\pm$ 0.0000020 \\
 $a$ [R$_\odot$]  & \multicolumn{2}{c}{7.37 $\pm$ 0.04 }        & 9.551 $\pm$ 0.026 \\
 $v_\gamma$ [km/s]& \multicolumn{2}{c}{-2.66 $\pm$ 0.08}        & -3.131 $\pm$ 0.089 \\
 $e$              & \multicolumn{2}{c}{--}                      & 0.190 $\pm$ 0.003   \\
 $\omega$ [deg]   & \multicolumn{2}{c}{--}                      & 161.83 $\pm$ 0.94   \\
 $q = M_2/M_1$    & \multicolumn{2}{c}{0.96 $\pm$ 0.02 }        & -- \\
 $i$ [deg]        & \multicolumn{2}{c}{85.17 $\pm$ 1.8 }        & -- \\
 $K$ [km/s]       &  101.9 $\pm$ 0.4  & 106.4 $\pm$ 0.4         & 40.40 $\pm$ 0.03 \\
 $T$ [K]          &  5170 (fixed)     & 5203 $\pm$ 75           & -- \\
 $M$ [M$_\odot$]  &  0.86 $\pm$ 0.02  & 0.83 $\pm$ 0.02         & -- \\
 $R$ [R$_\odot$]  &  1.23 $\pm$ 0.03  & 0.95 $\pm$ 0.02         & -- \\
 $M_{bol}$ [mag]  &  4.78 $\pm$ 0.05  & 5.32 $\pm$ 0.06         & -- \\
 $L_B [\%]$       &  16.5 $\pm$ 2.0   & 10.2 $\pm$ 2.0          & 73.3 $\pm$ 3.5 \\
 $L_V [\%]$       &  24.3 $\pm$ 2.3   & 14.9 $\pm$ 1.7          & 60.8 $\pm$ 4.0 \\
 $L_R [\%]$       &  26.3 $\pm$ 2.4   & 16.0 $\pm$ 1.6          & 57.7 $\pm$ 4.2 \\
 \hline
\end{tabular}}
\end{table}

\begin{table}
  \caption{The parameters of visual orbit of BR Ind.}
  \label{Table_orbit_BRInd}
  \centering
  \begin{tabular}{c c c c}
\hline \hline
Parameter       &  Our solution    & \cite{Seymour2002}  \\
 \hline
 $p_3$ [yr]     & 147.9 $\pm$ 2.5  & 167.0  \\
 $T_0$ [yr]     & 2050.3 $\pm$ 1.9 & 2056.0 \\
 $e$            & 0.711 $\pm$ 0.021& 0.521  \\
 $a$ [arcsec]   & 0.864 $\pm$ 0.045& 0.894  \\
 $i$ [deg]      & 154.6 $\pm$ 8.2  & 141.9  \\
 $\Omega$ [deg] & 220.8 $\pm$ 11.8 & 142.8  \\
 $\omega$ [deg] & 80.1 $\pm$ 9.9   & 178.4  \\
 \hline
\end{tabular}
\end{table}

The available positional measurements were analysed leading to slight improvement of the published
fit by \cite{Seymour2002}. The parameters are given in Table \ref{Table_orbit_BRInd} and the plot
of the orbit in Fig. \ref{FigBRInd_OC_orbit}. As one can see, the period is a bit shorter and the
eccentricity higher. The orientation of the orbit is really close to face-on ($i=155^\circ$), which
is in agreement with the discussion in the previous paragraph.

From the fit of the visual orbit we can also derive the total mass of the whole system. This
resulted in about 3.34~M$_\odot$. Such mass can be used to derive the individual masses of the
pair B in the system. If we accept the orbital solution as derived from our LC+RV analysis, the
mass of the pair B has to be of about 1.65~M$_\odot$. Due to the fact that the B subsystem is only
a SB1-type binary, we can only estimate its individual masses. For the SB1-type binary we can
calculate a so-called mass function \citep{2009ebs..book.....K}: $$f(M)_B = \frac{1}{2 \pi G} K^3
P_B \, (1-e_B^2)^{3/2} = \frac{(M_{Bb} \cdot \sin i_B)^3}{(M_{Ba}+M_{Bb})^2},$$ and from the
knowledge of the total mass of pair B we can derive the product $(M_{Bb} \cdot \sin i_B) =
0.47$~M$_\odot$. Obviously, we do not know the inclination of the pair, but at least some
estimation can be done also with these values. From the LC solution the third light of the pair B
is higher than the combined light coming from the two eclipsing components of pair A. Hence, this
finding is in excellent agreement with the masses as derived from the SB1 binary of the pair B and
the total mass -- but only with the assumption that the inclination is close to 90$^\circ$. Hence,
the two components of the B pair should be of about F8V+M2V. With such a configuration the
individual luminosity levels and their ratios as well as the non-detection of the Bb component in
the spectra will be explained. Also the spectral classification is now clarified, being of the Ba
component instead of the eclipsing pair. And finally, this solution would indicate that the two
visual components have their individual magnitudes shifted of about 1.1~mag, which is in agreement
with the magnitude differences as published in the WDS catalogue \citep{WDS}. However, the whole
discussion is solely based on the assumption that the Hipparcos parallax
\citep{2007A&A...474..653V} of 20.65~mas is correct.

\section{Discussion and Conclusions}

Regardless of the fact that a lot of work has been done on the theoretical modelling as well as on
the observations during the last decades, the formation of systems of higher multiplicity still
remains an open question. Multiplicity itself is the most promising mechanism to produce close
binaries with short orbital periods below 1 day. A third component may cause Kozai cycles and then
the tidal friction between the binary components causes an orbital shrinkage and its
circularization, see e.g. \cite{2001ApJ...562.1012E} and \cite{2016MNRAS.455.4136B}. Nevertheless,
this is not the only one possible scenario of origin of such systems and several other competing
theories are still being discussed. Truly existing systems were probably produced by a combination
of several different mechanisms, see e.g. \cite{2008MNRAS.389..925T}. The numerical simulations
which include only particular mechanisms are able to explain only some statistical properties of
the multiple systems, but fail to explain the others. That is a matter of intensive investigation
of recent years and each newly discovered multiple system with its known orbital and physical
parameters should help us to improve the statistical properties of the sample and provide us with
new observational constrains.

The study of three selected eclipsing multiple systems provides us with only a piece of information
needed for construction of the theory of stellar formation and evolution. Despite this fact, it is
still a valuable contribution to the topic in several aspects. At first, the three systems
represent the most typical multiple systems nowadays (containing the close inner pair with the
period of a few days and the distant component of much longer period). At second, their physical
and orbital properties are now well-determined and can be placed into the broader
context of theoretical modelling (e.g. the ratio of periods of the inner and outer orbits, the mass
ratios or the eccentricity values), see e.g. \cite{2003A&A...397..159H} or \cite{2008MNRAS.389..925T}.

At third, each of the studied systems is somehow interesting and deserves our attention. V773~Cas
was found to be much closer to the Sun than originally assumed on the basis of the Hipparcos data.
QS~Aql is rather massive, but moving around a common barycenter with the third component on the
orbit with a very high eccentricity of about 0.95. And also BR~Ind was found to be of a rare
quadruple system consisting of eclipsing and non-eclipsing pairs. This detection of the higher order multiplicity
among the stars of such a late type is also rather remarkable, because for the late type stars
their multiplicity fraction is generally very low, see e.g. \cite{2013ARA&A..51..269D}. As a common
characteristics for all these systems should also be mentioned that in all of them the most massive
and most luminous star is the distant third component (apart from BR~Ind, where it is a binary).
And as was already shown (e.g. by \citealt{2008MNRAS.389..925T} and \citealt{2016MNRAS.455.4136B}),
such systems are still rather rare.

As a by-product we also derived the possible mutual inclinations for these three systems and their
orbits. Due to the fact that longitude of the ascending node of the inner eclipsing pair is not
known, we can only estimate the ranges within which the mutual inclination should lie. This
resulted in 48--142~deg for V773~Cas, 22--145~deg for QS~Aql, and 69--120~deg for BR~Ind,
respectively. As one can see, the ranges are still rather high and the uncertainty should be lower
when knowing the longitude of the ascending node. However, this can only be achieved resolving the
inner eclipsing pair via interferometry, which is very problematic (due to the luminous third star
and a small angular separation of the eclipsing components). The most promising in this sense seems
to be V773~Cas, where the predicted angular distance was computed to be of about 0.8~mas.

And finally, such a study is also viable from the observational point of view. We can see that
there still exist many systems never analysed before and their parameters not known, despite the
fact that the observations exist and are easily obtainable. At this point it would be suitable to
mention that the photometric data for V773~Cas and QS~Aql were obtained using only the 34mm-sized
telescopes, while BR~Ind with a 107-mm one.

 \begin{acknowledgements}
We would like to thank the Pierre Auger Collaboration for the use of its facilities. The operation
of the robotic telescope FRAM is supported by the EU grant GLORIA (No. 283783 in FP7-Capacities
program) and by the grant of the Ministry of Education of the Czech Republic (MSMT-CR LM2015038).
The data calibration and analysis related to FRAM telescope is supported by the Ministry of
Education of the Czech Republic (MSMT-CR LG15014) and by the Czech Science Foundation grant No.
14-17501S. The observations obtained with the MPG 2.2m telescope were supported by the Ministry of
Education, Youth and Sports project - LG14013 (Tycho Brahe: Supporting Ground-based Astronomical
Observations). We would like to thank the observers S.~Ehlerov\'a, A.~Kawka, P.~Kab\'ath, and
S.~Vennes for obtaining the data. R.~K\v{r}\'{\i}\v{c}ek is also acknowledged for obtaining some of
the spectroscopic data. This research has made use of the Washington Double Star Catalog maintained
at the U.S. Naval Observatory.  This investigation was supported by the Czech Science Foundation
grants No. P209/10/0715 and GA15-02112S. We also do thank the {\sc ASAS}, and {\sc Pi of the sky}
teams for making all of the observations easily public available. This research has made use of the
SIMBAD and VIZIER databases, operated at CDS, Strasbourg, France and of NASA's Astrophysics Data
System Bibliographic Services.
 \end{acknowledgements}

%
%

\bibliographystyle{apj}
\bibliography{citace}

$\,$ \newpage


\begin{appendix} 

\section{Tables of minima}


\begin{table}
 \centering
\tiny
 \caption{List of new minima timings used for the analysis} \label{minima}
\begin{tabular}{ccclcl}
\hline\hline\noalign{\smallskip}
 Star       &    JD Hel.- &  Error & Type   &  Filter  & Source /     \\
            &   2400000   &  [day] &        &          & Observatory  \\
\noalign{\smallskip}\hline \noalign{\smallskip}
 V773 Cas & 48500.8874  & 0.0095  & Prim & Hp  & Hipparcos \\
 V773 Cas & 54798.48508 & 0.00154 & Prim & BVR & PS  \\
 V773 Cas & 55062.39534 & 0.00139 & Prim & BVR & PS  \\
 V773 Cas & 55071.45118 & 0.00133 & Prim & BVR & PS  \\
 V773 Cas & 55410.39017 & 0.00197 & Prim & BVR & PS  \\
 V773 Cas & 55419.44651 & 0.00101 & Prim & BVR & PS  \\
 V773 Cas & 55481.54064 & 0.00182 & Prim & BVR & PS  \\
 V773 Cas & 55754.50689 & 0.00038 & Sec  & I   & RU  \\
 V773 Cas & 55776.49780 & 0.00026 & Prim & R   & PS  \\
 V773 Cas & 55776.49748 & 0.00066 & Prim & I   & RU  \\
 V773 Cas & 55877.40172 & 0.00047 & Prim & R   & RU  \\
 V773 Cas & 56155.54350 & 0.00069 & Sec  & R   & RU  \\
 V773 Cas & 56159.42162 & 0.00105 & Prim & R   & RU  \\
 V773 Cas & 56230.57514 & 0.00062 & Sec  & C   & RU  \\
 V773 Cas & 56252.56717 & 0.00075 & Prim & C   & RU  \\
 V773 Cas & 56304.31383 & 0.00040 & Prim & C   & RU  \\
 V773 Cas & 56516.47530 & 0.00062 & Prim & I   & RU  \\
 V773 Cas & 56538.46837 & 0.00084 & Sec  & C   & RU  \\
 V773 Cas & 56851.53476 & 0.00029 & Sec  & R   & RU  \\
 V773 Cas & 56930.44650 & 0.00031 & Prim & I   & RU  \\
 V773 Cas & 56978.31353 & 0.00019 & Sec  & I   & RU  \\
 V773 Cas & 57199.52681 & 0.00057 & Prim & R   & RU  \\
 V773 Cas & 57234.46016 & 0.00090 & Sec  & R   & RU  \\
 V773 Cas & 57348.30130 & 0.00075 & Sec  & R   & RU  \\
 V773 Cas & 57383.23313 & 0.00048 & Prim & R   & RU  \\
 V773 Cas & 57569.51959 & 0.00035 & Prim & R   & RU  \\
  \hline
 QS Aql   & 55063.40382 & 0.00052 & Prim & R   & PS  \\
 QS Aql   & 55068.43043 & 0.00350 & Prim & BVRI& RU  \\
 QS Aql   & 55068.43052 & 0.00032 & Prim & R   & PS  \\
 QS Aql   & 55102.36336 & 0.00351 & Sec  & I   & RU  \\
 QS Aql   & 55405.20687 & 0.00170 & Prim & BVRI& RU  \\
 QS Aql   & 55480.60724 & 0.00241 & Prim & BVRI& RU  \\
 QS Aql   & 55817.39258 & 0.00155 & Prim & C   & RU  \\
 QS Aql   & 56112.70283 & 0.00113 & Sec  & R   & RU  \\
 QS Aql   & 56113.95940 & 0.00350 & Prim & R   & RU  \\
 QS Aql   & 56459.54374 & 0.01260 & Sec  & C   & RU  \\
 QS Aql   & 56488.44384 & 0.00122 & Prim & C   & RU  \\
 QS Aql   & 56870.47101 & 0.00045 & Prim & V   & RU  \\
 QS Aql   & 57242.43237 & 0.00136 & Prim & V   & RU  \\
 QS Aql   & 57531.46696 & 0.00212 & Prim & V   & RU  \\
 QS Aql   & 57580.47669 & 0.00195 & Sec  & V   & RU  \\
 QS Aql   & 57614.40355 & 0.00116 & Prim & V   & RU  \\
   \hline
 BR Ind   & 48512.97695 & 0.00227 & Prim & Hp  & Hipparcos \\
 BR Ind   & 48513.87160 & 0.00289 & Sec  & Hp  & Hipparcos \\
 BR Ind   & 52717.96502 & 0.00264 & Prim & V   & ASAS \\
 BR Ind   & 52718.86393 & 0.00171 & Sec  & V   & ASAS \\
 BR Ind   & 53537.54375 & 0.00136 & Prim & V   & ASAS \\
 BR Ind   & 53538.43800 & 0.00197 & Sec  & V   & ASAS \\
 BR Ind   & 54628.52452 & 0.00203 & Prim & V   & ASAS \\
 BR Ind   & 54629.41872 & 0.00287 & Sec  & V   & ASAS \\
 BR Ind   & 56849.76183 & 0.00437 & Prim & I   & FRAM \\
 BR Ind   & 56857.79538 & 0.00186 & Sec  & I   & FRAM \\
 BR Ind   & 56940.82231 & 0.00777 & Prim & I   & FRAM \\
 BR Ind   & 56941.72129 & 0.02155 & Sec  & I   & FRAM \\
 BR Ind   & 56857.79712 & 0.00155 & Sec  & I   & FRAM \\
 BR Ind   & 56941.72189 & 0.00119 & Sec  & I   & FRAM \\
 BR Ind   & 57189.91885 & 0.00013 & Sec  & B   & FRAM \\
 BR Ind   & 57264.91192 & 0.00720 & Sec  & B   & FRAM \\
 BR Ind   & 57272.94910 & 0.00060 & Prim & B   & FRAM \\
 BR Ind   & 57273.83371 & 0.00437 & Sec  & B   & FRAM \\
 BR Ind   & 57274.72655 & 0.00397 & Prim & B   & FRAM \\
 BR Ind   & 57282.75891 & 0.00281 & Sec  & B   & FRAM \\
 BR Ind   & 57283.64606 & 0.00143 & Prim & B   & FRAM \\
 BR Ind   & 57325.61535 & 0.01125 & Sec  & B   & FRAM \\
 BR Ind   & 57326.49662 & 0.00326 & Prim & B   & FRAM \\
 \noalign{\smallskip}\hline
\end{tabular}\\
Note: This table is available in its entirety in machine-readable and Virtual Observatory (VO)
forms. RU and PS are initials of the co-authors names.
\end{table}


\section{Tables of radial velocities}

\begin{table}
 \centering
 \tiny \caption{List of the radial velocities used for the analysis.} \label{RVs}
\begin{tabular}{ccrrc}
\hline\hline\noalign{\smallskip}
 Star             &  JD Hel.- & $RV_1$      & $RV_2$      & Observatory \\
                  & 2400000   &[km s$^{-1}$]&[km s$^{-1}$]&             \\
\noalign{\smallskip}\hline \noalign{\smallskip}
 V773 Cas & 56534.5191 &   22.964  &  -6.638  & Ond \\
 V773 Cas & 56540.5417 &  -82.834  &  96.039  & Ond \\
 V773 Cas & 56563.5964 &  -85.927  & 107.335  & Ond \\
 V773 Cas & 56572.5618 &   94.114  & -81.845  & Ond \\
 V773 Cas & 56590.3591 &   43.175  & -12.803  & Ond \\
 V773 Cas & 56596.5316 &   42.319  & -24.662  & Ond \\
 V773 Cas & 56665.2559 &    9.844  &  13.742  & Ond \\
 V773 Cas & 56666.2016 &   74.693  & -71.150  & Ond \\
 V773 Cas & 56862.4161 &   97.931  & -83.252  & Ond \\
 V773 Cas & 56862.5705 &   99.085  & -83.225  & Ond \\
 V773 Cas & 56924.3821 &   87.160  & -78.330  & Ond \\
 V773 Cas & 57260.5698 &   57.533  & -50.438  & Ond \\
 V773 Cas & 57275.4178 &  -72.467  &  91.545  & Ond \\
 V773 Cas & 57284.4024 &  101.586  & -89.144  & Ond \\
 V773 Cas & 57287.4956 &    8.957  &   7.953  & Ond \\
 V773 Cas & 55497.4625 &  -71.760  &  84.909  & Ond \\
 V773 Cas & 55622.3217 &  -44.576  &  59.889  & Ond \\
 V773 Cas & 55623.3745 &   93.110  & -84.606  & Ond \\
 V773 Cas & 56192.3778 &  103.565  & -89.690  & Ond \\
 V773 Cas & 56400.5898 &  -91.459  &  92.108  & Ond \\  \hline
 QS Aql   & 56013.6415 &  -47.198  &  66.622  & Ond \\
 QS Aql   & 56041.5475 &  -80.691  & 175.642  & Ond \\
 QS Aql   & 56151.4539 &   19.518  &-107.258  & Ond \\
 QS Aql   & 56192.3915 &  -91.426  & 156.803  & Ond \\
 QS Aql   & 56357.6687 &   -1.965  & -64.234  & Ond \\
 QS Aql   & 56400.5732 &  -42.711  &  41.075  & Ond \\
 QS Aql   & 56407.5370 &   55.993  &-207.647  & Ond \\
 QS Aql   & 56464.4176 &  -46.862  &  61.639  & Ond \\
 QS Aql   & 56527.4023 &  -12.191  &    --    & Ond \\
 QS Aql   & 56534.3037 &  -90.273  & 172.925  & Ond \\
 QS Aql   & 56534.5020 &  -83.692  & 167.165  & Ond \\
 QS Aql   & 56540.2671 &   32.763  &-140.315  & Ond \\
 QS Aql   & 56590.3185 &   -0.375  & -79.928  & Ond \\
 QS Aql   & 56665.1783 &  -79.765  & 165.063  & Ond \\
 QS Aql   & 56736.6346 &   59.747  &-223.658  & Ond \\
 QS Aql   & 56772.5975 &  -45.892  &  63.491  & Ond \\
 QS Aql   & 56799.4995 &   57.337  &-218.136  & Ond \\
 QS Aql   & 56862.4011 &   60.692  &-204.877  & Ond \\
 QS Aql   & 56924.3353 &  -43.001  &  63.641  & Ond \\
 QS Aql   & 57126.5445 &   32.494  &-142.932  & Ond \\
 QS Aql   & 57208.5380 &   -5.866  &    --    & Ond \\
 QS Aql   & 57211.3821 &   39.267  &-167.650  & Ond \\
 QS Aql   & 57245.4581 &  -88.122  & 169.952  & Ond \\
 QS Aql   & 57260.4073 &  -77.626  & 156.522  & Ond \\
 QS Aql   & 57275.3864 &  -59.986  & 112.408  & Ond \\
 QS Aql   & 57294.3430 &   42.652  &-156.705  & Ond \\
 QS Aql   & 57323.3276 &  -79.242  & 153.787  & Ond \\
 QS Aql   & 57330.2758 &    4.005  & -79.303  & Ond \\
 QS Aql   & 56555.5332 &   54.385  &-217.627  & FEROS \\
 QS Aql   & 56555.5738 &   58.308  &-222.318  & FEROS \\
 QS Aql   & 56555.6103 &   59.994  &-222.843  & FEROS \\
 QS Aql   & 56555.6476 &   59.324  &-231.485  & FEROS \\
 QS Aql   & 56562.5281 &  -25.677  &  19.208  & FEROS \\
 QS Aql   & 56562.5807 &  -16.597  &  -2.292  & FEROS \\
 QS Aql   & 56563.5079 &   36.519  &-171.487  & FEROS \\
 QS Aql   & 56563.5441 &   29.523  &-163.707  & FEROS \\
 QS Aql   & 56563.6067 &   25.396  &-141.287  & FEROS \\
 QS Aql   & 56564.5231 &  -89.022  & 187.011  & FEROS \\
 QS Aql   & 56564.5565 &  -86.879  & 189.186  & FEROS \\
 QS Aql   & 56564.5972 &  -86.505  & 180.737  & FEROS \\  \hline
 BR Ind   & 56555.5550 & -103.332  & 106.265  & FEROS \\
 BR Ind   & 56555.5951 & -105.425  & 105.437  & FEROS \\
 BR Ind   & 56555.6312 & -104.541  & 103.410  & FEROS \\
 BR Ind   & 56555.6801 &  -98.914  &  96.941  & FEROS \\
 BR Ind   & 56555.7187 &  -91.928  &  89.599  & FEROS \\
 BR Ind   & 56555.7449 &  -89.879  &  86.619  & FEROS \\
 BR Ind   & 56555.7783 &  -81.036  &  83.418  & FEROS \\
 BR Ind   & 56555.8036 &  -76.804  &  76.149  & FEROS \\
 BR Ind   & 56562.5654 &  -87.647  &  87.267  & FEROS \\
 BR Ind   & 56562.6090 &  -98.574  &  95.403  & FEROS \\
 BR Ind   & 56562.6492 &  -96.409  & 104.060  & FEROS \\
 BR Ind   & 56562.6864 & -104.150  & 102.517  & FEROS \\
 BR Ind   & 56562.7140 & -101.171  & 104.689  & FEROS \\
 BR Ind   & 56562.7327 & -108.033  & 106.827  & FEROS \\
 BR Ind   & 56562.7573 & -103.429  & 101.504  & FEROS \\
 BR Ind   & 56563.5972 &   95.514  &-109.126  & FEROS \\
 BR Ind   & 56563.6346 &  100.678  &-109.995  & FEROS \\
 BR Ind   & 56563.6621 &   95.935  &-109.043  & FEROS \\
 BR Ind   & 56563.6965 &   91.273  &-105.220  & FEROS \\
 BR Ind   & 56563.7209 &   89.127  &-102.994  & FEROS \\
 BR Ind   & 56563.7401 &   91.691  &-100.083  & FEROS \\
 BR Ind   & 56563.7676 &   87.438  & -91.302  & FEROS \\
 BR Ind   & 56564.5380 & -104.525  &  95.091  & FEROS \\
 BR Ind   & 56564.5664 & -102.728  &  95.506  & FEROS \\
 BR Ind   & 56564.6411 &  -98.061  &  93.281  & FEROS \\
 BR Ind   & 56564.6799 &  -89.713  &  87.975  & FEROS \\
 BR Ind   & 56564.7262 &  -86.142  &  80.363  & FEROS \\
 BR Ind   & 56986.5334 &   60.691  & -60.939  & FEROS \\
 BR Ind   & 56989.5235 & -105.452  & 100.628  & FEROS \\
 BR Ind   & 56990.5286 &   86.391  & -91.730  & FEROS \\
 BR Ind   & 56994.5642 &  -59.478  &  57.979  & FEROS \\
 BR Ind   & 57153.9298 &  -83.130  &  83.548  & FEROS \\
 BR Ind   & 57155.9194 &  -25.281  &          & FEROS \\
 BR Ind   & 57157.9286 &   54.533  & -67.251  & FEROS \\
 BR Ind   & 57159.9232 &   90.127  &-101.785  & FEROS \\
 BR Ind   & 57312.5675 & -100.841  &  99.968  & FEROS \\
 BR Ind   & 57598.7205 &  -27.285  &  33.544  & FEROS \\
 BR Ind   & 57599.7549 &  -25.503  &          & FEROS \\
 BR Ind   & 57600.7793 &   64.952  & -70.670  & FEROS \\
 BR Ind   & 57602.7507 &   97.304  &-107.223  & FEROS \\
\noalign{\smallskip}\hline
\end{tabular}\\
Note: This table is available in its entirety in machine-readable and Virtual Observatory (VO)
forms.
\end{table}

\end{appendix}
\end{document}